\documentclass[letter]{aa} 
\usepackage{graphicx}
\usepackage{natbib}
\usepackage{txfonts}

\begin{document}

\title{Deep near-infrared interferometric search for low-mass companions around $\beta$~Pictoris\thanks{Based on observations collected at the ESO La Silla Paranal Observatory under program IDs 084.C-0566 and 384.C-0806.}}

\titlerunning{Near-infrared interferometric search for companions around $\beta$~Pic}

\author{O. Absil \inst{1}\fnmsep\thanks{FNRS Postdoctoral Researcher, email: \texttt{absil@astro.ulg.ac.be}}
\and J.-B. Le Bouquin \inst{2}
\and J.~Lebreton \inst{2}
\and J.-C.~Augereau \inst{2}
\and M.~Benisty \inst{3}
\and G.~Chauvin \inst{2}
\and C.~Hanot \inst{1}
\and \\ A.~M{\'e}rand \inst{4}
\and G.~Montagnier \inst{4}}

\institute{Institut d'Astrophysique et de G\'eophysique, Universit\'e de Li\`ege, 17 All\'ee du Six Ao\^ut, B-4000 Li\`ege, Belgium
\and
LAOG--UMR 5571, CNRS and Universit\'e Joseph Fourier, BP 53, F-38041 Grenoble, France
\and
INAF - Osservatorio Astrofisico di Arcetri, Largo E. Fermi 5, 50125 Firenze, Italy
\and
European Southern Observatory, Casilla 19001, Santiago 19, Chile}

\date{Received 4 June 2010; accepted 3 September 2010}


\abstract
{}
{We search for low-mass companions in the innermost region ($<300$~mas, i.e., 6~AU) of the $\beta$~Pic planetary system.}
{We obtained interferometric closure phase measurements in the K-band with the VLTI/AMBER instrument used in its medium spectral resolution mode. Fringe stabilization was provided by the FINITO fringe tracker.}
{In a search region of between 2 and 60~mas in radius, our observations exclude at $3\sigma$ significance the presence of companions with K-band contrasts greater than $5\times 10^{-3}$ for 90\% of the possible positions in the search zone (i.e., 90\% completeness). The median $1\sigma$ error bar in the contrast of potential companions within our search region is $1.2\times 10^{-3}$. The best fit to our data set using a binary model is found for a faint companion located at about 14.4~mas from $\beta$~Pic, which has a contrast of $1.8\times 10^{-3} \pm 1.1\times 10^{-3}$ (a result consistent with the absence of companions). For angular separations larger than 60~mas, both time smearing and field-of-view limitations reduce the sensitivity.}
{We can exclude the presence of brown dwarfs with masses higher than $29\,M_{\rm Jup}$ (resp.\ $47\,M_{\rm Jup}$) at a 50\% (resp.\ 90\%) completeness level within the first few AUs around $\beta$~Pic. Interferometric closure phases offer a promising way to directly image low-mass companions in the close environment of nearby young stars.}


\maketitle


\section{Introduction}

The young \citep[$\sim 12$~Myr,][]{Zuckerman01}, nearby (19.3~pc), and bright ($K=3.5$) A5V-type star $\beta$~Pictoris (HD~39060) is surrounded by one of the most famous extrasolar planetary systems, consisting of a recently detected planetary companion \citep{Lagrange09a,Lagrange10} inside an optically thin debris disk seen edge-on \citep{Smith84}, which has been resolved at various wavelengths. Several asymmetries have been identified in the debris disk, including a warp at $\sim 50$~AU \citep{Heap00} that is now understood to be the result of the dynamical influence of a massive body (a few Jupiter masses) on an eccentric orbit around the central star \citep{Freistetter07}. The 9-$M_{\rm Jup}$ companion discovered by \citet{Lagrange09a} may be the cause of this warp. We note that the planetary nature of this companion was not easy to ascertain \citep{Lagrange09b}, because the companion was located at a projected distance smaller than the inner working angle of VLT/NACO (335~mas for $9\,M_{\rm Jup}$) between the discovery observations in 2003 and the confirmation observations in late 2009.

Long-baseline optical interferometry is a promising technique to search for faint companions at angular separations smaller than the diffraction limit of a single aperture. In particular, closure phase measurements on a closed triangle of baselines are very sensitive to asymmetries in the brightness distribution of the source, and can be used to detect faint companions. Closure phases have the added advantage of being insensitive to telescope-specific phase errors (unlike visibilities and phases), including atmospheric turbulence effects \citep[for a review of closure phases, see][]{Monnier03}.

The interferometric detection of extrasolar planets (hot Jupiters in particular) has already been attempted by a few groups, using in particular precision closure phase measurements. Despite the exquisite accuracy that has already been reached \citep[e.g., $0\fdg1$ stability in the CHARA/MIRC closure phases,][]{Zhao08}, no extrasolar planet has yet been detected. Differential phase techniques have not been more successful because of atmospheric and instrumental limitations \citep[e.g.,][]{Millour08,Matter10}. Higher sensitivities to faint companions can be reached when closure phases are obtained on fully resolved stellar photospheres \citep[e.g.,][]{Lacour08,Duvert10}, but this is unfortunately not the case for most main-sequence stars with currently available interferometric baselines. In this Letter, we perform a deep interferometric search for faint companions at short angular distances from the unresolved young main sequence star $\beta$~Pic \citep[$\theta_{\rm LD}=0.85\pm0.12$~mas,][]{DiFolco04}, based on closure phase measurements with VLTI/AMBER.


\section{Observations and data reduction} \label{sec:obs}

Observations of $\beta$~Pic were performed on four different nights from 2010 January 24 to 28 with the AMBER instrument, used in its medium resolution mode ($R\simeq1500$ from 1.93 to 2.27~$\mu$m) with three 1.8-m Auxiliary Telescopes of the VLTI arranged on the A0-G1-K0 triangle (ground baselines from 90~m to 128~m). Fringe tracking was provided by the FINITO facility during all our observations. A total of 12~Observing Blocks (OBs) were completed for $\beta$~Pic, with Detector Integration Times (DIT) ranging from 0.5 to 1~sec depending on the atmospheric conditions. To reduce time smearing in our data set, long OBs were divided into shorter sequence of 15~min or less, giving a total of 26~OBs. Our observations of $\beta$~Pic were interleaved with observations of HD~39640, a G8III calibration star of magnitude $K=3.0$ and angular diameter $\theta_{\rm LD}=1.235\pm0.015$~mas \citep{Merand05}, located only $1\fdg2$ from $\beta$~Pic on the sky.

\begin{figure}[t]
\centering
\includegraphics[width=9cm]{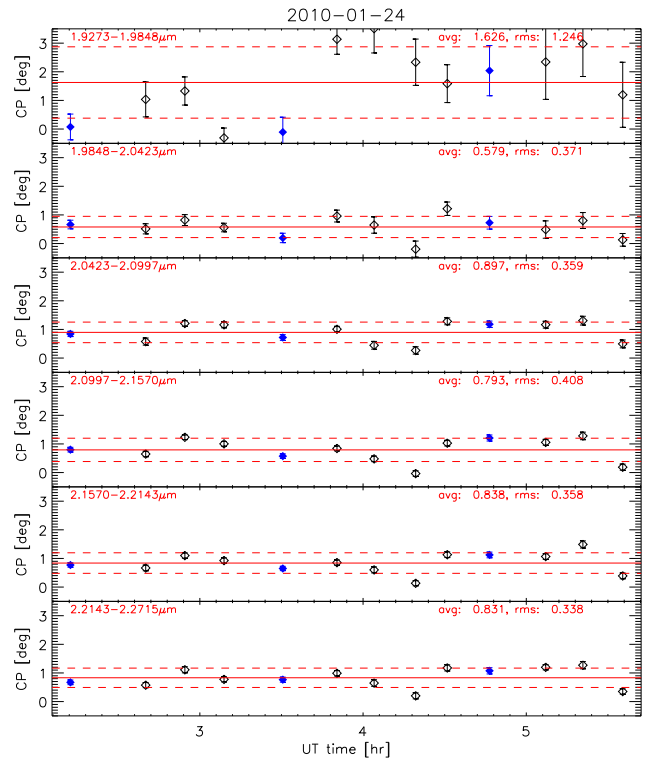}
\caption{Fluctuation of the closure phase as a function of time for $\beta$~Pic (black empty diamonds) and its calibrator (blue filled diamonds) during the night of January 24. Data have been binned into 6 spectral channels to reduce statistical errors. The solid lines represent the mean of the CP transfer functions and the dashed lines their standard deviations.}
\label{fig:CPtransfunc}
\end{figure}

Raw data were reduced using the amdlib v3.0 package \citep{Tatulli07,Chelli09}, using all recorded detector frames (no fringe selection). The $\beta$~Pic closure phases were calibrated by a simple subtraction of the average closure phase of all calibrator measurements obtained during the same night with the same instrumental set-up. This procedure implicitly assumes that the closure phase transfer function does not vary significantly during several hours of observations, a hypothesis that we verified during our four observing nights. The error bar related to the calibration process is generally estimated by measuring the standard deviation in the calibrator closure phases during the night. In our case, too few calibrator measurements are available to provide a reliable error bar, so we decided to use the standard deviation in all measurements (science and calibrator) instead. It must be noted that, if the scientific target were to have a companion producing a significant closure phase signal, this would lead to an overestimated calibration error bar and could possibly hide the companion (we address this in the last paragraph of Sect.~\ref{sub:analysis}). The evaluation of the calibration error bar is illustrated in Fig.~\ref{fig:CPtransfunc} for the night of January 24, which is actually the poorest night in our data set in terms of transfer function stability. In this plot, the data were binned into 6 spectral channels to reduce their statistical errors and isolate the effect of transfer function instabilities. The short end of the K band (top panel of Fig.~\ref{fig:CPtransfunc}) displays significant instabilities in its closure phases, most probably caused by strong and variable spectral lines in the Earth atmosphere. These instabilities disappear beyond about $2.0~\mu$m, so we restrict our spectral range to the 2.00--2.26~$\mu$m region. In this spectral range, we find that the RMS calibration errors range between $0\fdg20$ and $0\fdg37$ depending on the night. These error bars will be added quadratically for our whole data set, night by night (after binning individual spectral channels, where required).  

\section{Searching for faint off-axis companions}

    \subsection{Data analysis} \label{sub:analysis}

The field-of-view (FOV) of the AMBER instrument is limited by the use of single-mode fibers. We estimated the off-axis transmission of a point-like object in the presence of atmospheric turbulence corrected for tip-tilt fluctuations by the STRAP system on the ATs. Our simulations produced a Gaussian transmission with a full width at half maximum of 420~mas for the median seeing of our observations ($\sim 0\farcs8$). For angular distances larger than 210~mas, the transmission quickly drops to low values and we limit our analysis to separations smaller than about 300~mas, where the transmission drops by a factor of~4. This corresponds to a linear distance of $\sim6$~AU from $\beta$~Pic. We note however that, because of the Earth's rotation, closure phases are variable in time. In order not to smear the closure phase signal too much during our $\sim$15~min OBs, we had to restrict our search region to a FOV radius of about 50~mas. A finer time sampling would be needed to explore the full 300~mas FOV at the highest sensitivity.


The first step in our data analysis is to adapt the spectral resolution to the explored FOV, by ensuring that the variations in the closure phase as a function of wavelength, created by a potential companion, are sampled with at least four data points per period. The periodicity in the closure phase signal as a function of wavelength is given roughly by $P_{\lambda} = \lambda^2/(B\theta-\lambda)$, where $B$ is the mean interferometric baseline length. For $B\simeq100$~m in our case, and a maximum angular separation of 50~mas, this implies that $P_{\lambda} \simeq 0.2~\mu$m. The appropriate spectral bin size to cover a 50~mas FOV is thus $\Delta\lambda=P_{\lambda}/4=50$~nm, so that a total of 5 spectral channels are needed across the $2.00-2.26~\mu$m spectral range. We used a 5-sigma clipping method to remove outliers when binning our data into the synthetic spectral channels, and statistical error bars are estimated to be the standard deviation in the individual data points of each bin.

\begin{figure*}[t]
\centering
\includegraphics[width=17cm]{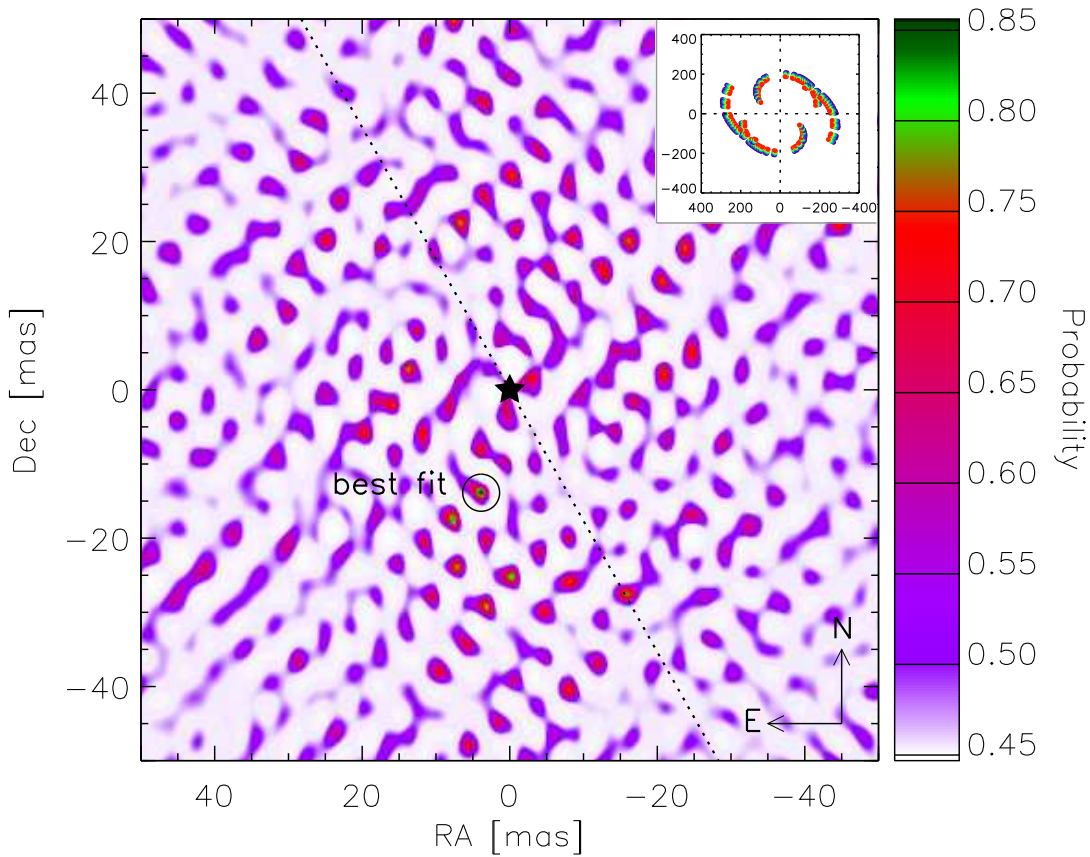}
\caption{Probability of a binary model to reproduce our data set, for various positions of the secondary companion across a FOV 50~mas in radius. At all places in the FOV, the flux of the companion has been tuned to minimize the $\chi_r^2$. The zones in white have their minimum $\chi_r^2$ for a companion of zero flux. The position of $\beta$~Pic is represented by a star, and the orientation of the circumstellar disk midplane \citep[${\rm PA}=29\fdg5$,][]{Boccaletti09} by a black dotted line. The $u,v$ coverage of our data set is represented in the upper right inset, including wavelength dependence (using colors from blue to red), with the axes graduated in cycles per arcsec.}
\label{fig:chi2maps}
\end{figure*}

The next step is to search the entire FOV for possible companions. To do this, we used the photospheric model of $\beta$~Pic proposed by \citet{DiFolco04}, to which we added a secondary point-like companion at various locations within the FOV. The $\chi^2$ distance between the data and all our binary models was then computed, and for each position we selected the companion contrast that produces the lowest $\chi^2$. The resulting $\chi^2$ map was then converted into a probability map (Fig.~\ref{fig:chi2maps}) using the $\chi^2$ probability distribution function, taking into account the number of degrees of freedom in our $\chi^2$ distribution. In the present case, there are $26\times5$ independent data points and 1 parameter to fit (the companion contrast for each binary position), hence 129 degrees of freedom. In the absence of a companion, the reduced chi square ($\chi_r^2$) amounts to 1.01, which corresponds to a probability of 45\% for our photospheric model alone to reproduce the data set.

\begin{figure}[t]
\centering
\includegraphics[width=9cm]{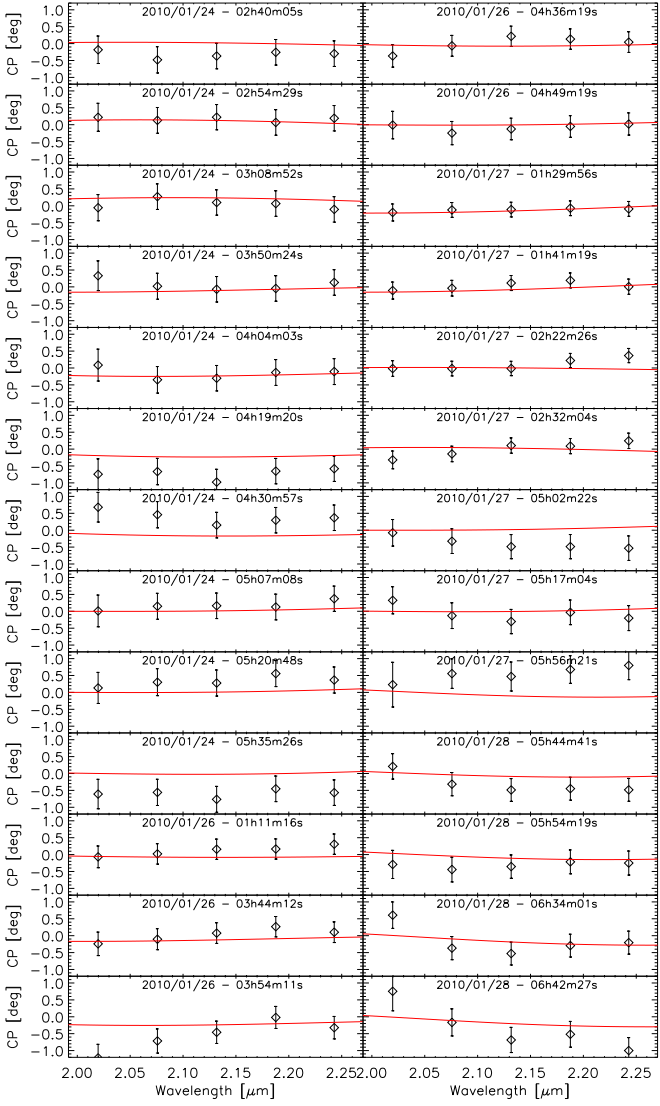}
\caption{Representation of the whole data set (diamonds with error bars, binned into 5 spectral channels) and of the best-fit model (red curves) found on the 50~mas radius FOV. This model corresponds to a companion with a contrast of $1.8\times 10^{-3}$ located at cartesian coordinates $\Delta{\rm RA}=3.9$~mas and $\Delta{\rm Dec}=-13.9$~mas relative to the central star.}
\label{fig:data}
\end{figure}

Our probability map shows a local maximum of 86\% (i.e., $\chi_r^2=0.87$) at a position $(\Delta{\rm RA},\,\Delta{\rm Dec})=(3.9,\,-13.9)$~mas, with a best-fit contrast of $1.8\times 10^{-3}$. The quality of our best-fit model, illustrated in Fig.~\ref{fig:data}, confirms that this model closely reproduces our data set. Several other companion positions within our FOV would provide (almost) equally good fits. We note however that the pure photospheric model also reproduces the data quite well, so that there is no real evidence of a companion. We now verify whether this could be caused by an overestimation of the calibration error bars, as discussed in Sect.~\ref{sec:obs}. We artificially decrease our estimated calibration error bars until we achieve $\chi^2_r=1$ for our best-fit model (which is reached when calibration error bars are multiplied by 0.92). The significance of this possible detection can then be estimated by converting the probability of the null hypothesis (i.e., no companion), which is now only 9\%, into an equivalent number of standard deviations for the possible detection of a companion, using standard relationships for normal distributions. We infer a $1.7\sigma$ significance for our detection, giving a companion contrast of $1.8\times 10^{-3} \pm 1.1\times 10^{-3}$. This low significance confirms that it cannot be considered as a real detection.

    \subsection{Sensitivity limits}

Sensitivity limits can be derived from our $\chi_r^2$ map (or equivalently, probability map), by searching at each point of the FOV for the companion contrast that would produce a $\chi_r^2$ larger than a pre-defined threshold. We choose a $3\sigma$ criterion to define our sensitivity limit (i.e., probability of less than 0.27\% for the model to reproduce the data). With 129 degrees of freedom, this corresponds to a threshold $\chi^2_{r,{\rm lim}} = 1.38$. Once the $3\sigma$ upper limit to the detectable companion contrast is computed at each point of the FOV, one can define global sensitivity limits by building and inspecting the histogram of the $3\sigma$ detection levels across the considered FOV. For instance, for the 50~mas FOV used in Fig.~\ref{fig:chi2maps}, the contrast upper limit is $<3.6\times 10^{-3}$ for 50\% of all positions in the FOV ({\em 50\% completeness}) and $<4.8\times 10^{-3}$ for 90\% completeness. This confirms that the typical $1\sigma$ error bar in the contrast of a detected companion would be about $1.2\times 10^{-3}$, as estimated in the previous paragraph. 

These sensitivity limits were confirmed by a double-blind test, where synthetic companions of various contrasts were introduced into our raw data set. In these blind tests, we were able to retrieve the companions with a contrast of $3.0\times 10^{-3}$ in about 50\% of the cases (although with a formal significance generally between 2 and $3\sigma$), and the companions with a contrast of $5.0\times 10^{-3}$ in all cases. These tests confirm the validity of our contrast upper limits based on the $\chi_r^2$ analysis.

\begin{figure}[t]
\centering
\includegraphics[width=9cm]{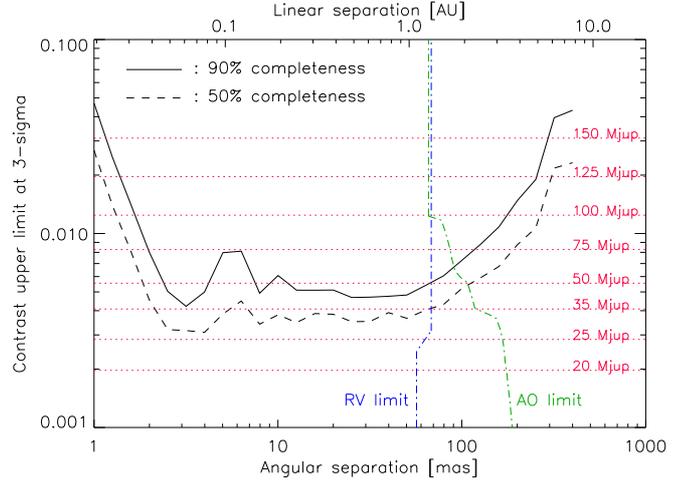}
\caption{Sensitivity curves showing the $3\sigma$ upper limit to the contrast of off-axis companions as a function of the angular separation for 50\% and 90\% completeness, computed across annular fields-of-view with 10\% relative width. Equivalent masses were computed using the COND model of \citet{Baraffe03}, for an age of 12~Myr. The companion discovery zones of radial velocity measurements \citep[][left of the blue dash-dotted line]{Galland06} and of AO-assisted coronagraphic imaging \citep[][right of the green dash-dotted line]{Boccaletti09} are shown for comparison.}
\label{fig:sensitivity}
\end{figure}

Sensitivity limits can also be computed as a function of angular separation, by building $\chi_r^2$ maps on annular fields-of-view of increasing size. This is illustrated in Fig.~\ref{fig:sensitivity}, where annuli 10\% in relative width have been used. This figure shows that VLTI/AMBER reaches its optimum sensitivity in the 2--60~mas region, where a median contrast of $3.5\times 10^{-3}$ (\ $5.0\times 10^{-3}$) can be reached at a 50\% (90\%) completeness level using the ATs. Beyond 60~mas, the effect of time smearing on the closure phase signal of potential companions becomes significant, reducing the sensitivity. For larger separations ($>200$~mas), the sensitivity degrades more rapidly because of the decreasing off-axis transmission of the single-mode fibers. The inner working angle of our interferometric study is about 1~mas, where the sensitivity drops to a few percent in contrast (e.g., $5\times10^{-2}$ at 90\% completeness) because of the limited angular resolution provided by our baselines.

\section{Discussion}

To compare our sensitivity limits with other studies, it is useful to express them in terms of companion masses. The first step is to convert contrasts into absolute magnitudes, taking into account the K magnitude and distance of $\beta$~Pic. Absolute magnitudes are then converted into masses using the COND evolutionary models of \citet{Baraffe03}, assuming an age of 12~Myr. The result is illustrated in Fig.~\ref{fig:sensitivity}, where mass upper limits are represented by dotted lines. In the 2--60~mas region, the median upper limit to companion masses is $29\,M_{\rm Jup}$ ($47\,M_{\rm Jup}$) for 50\% (90\%) completeness.

These upper limits should be compared with the main planet search methods around nearby main-sequence stars: radial velocity (RV) monitoring and direct (single-pupil) imaging. State-of-the-art RV measurements obtained with HARPS on $\beta$~Pic \citep{Galland06} have reached an accuracy of 180~m\,s$^{-1}$ (after correction for its pulsations), which provides a typical sensitivity of $10~M_{\rm Jup}$ for a semi-major axis $a=1$~AU. The sensitivity then scales as $a^{-2}$. The largest semi-major axis that can be reached depends on the timescale of the RV monitoring: to cover most of our interferometric search region (up to 6~AU in semi-major axis), $\beta$~Pic should be surveyed for about 5~years. For this time coverage, the only companions that could be detected by our interferometric search and not by the RV monitoring would be those located at orbital distances larger than 6~AU, which would by chance be at projected angular separations smaller than 300~mas at the time of our observations. K-band direct imaging observations using a Four Quadrant Phase Mask (FQPM) coronagraph on VLT/NACO have yielded an upper limit of $2.5\times 10^{-4}$ to the contrast of off-axis companions at projected distances $>335$~mas \citep{Lagrange09b}, and can reach a contrast of $1.3\times 10^{-2}$ at the FQPM inner working angle of 70~mas \citep{Boccaletti09}. In practice, the NACO-FQPM sensitivity becomes superior to that of AMBER for angular distances larger than $\sim100$~mas, so the two techniques can be considered complementary.


Although our detection limits are promising and would allow brown dwarfs to be detected around bright young stars such as $\beta$~Pic, they are insufficient to detect the planetary companion discovered by \citet{Lagrange09a}, which has an estimated K-band contrast of $2.5\times10^{-4}$. The accuracy of our measurements needs to be improved by at least a factor of 10 to reach this level of performance. The same improvement in measurement accuracy would be required to bring our sensitivity down to the realm of hot-Jupiter type planets. It is expected that the next generation of interferometric instruments at the VLTI (PIONIER, GRAVITY, and eventually VSI) could provide the necessary improvements in instrumental stability and sensitivity to achieve this performance.

\begin{acknowledgements}
The authors thank the anonymous referee for a thoughtful report that improved the data analysis. O.A.\ and C.H.\ acknowledge support from the ``Communaut\'e Fran\c caise de Belgique -- Actions de Recherche Concert\'ees -- Acad\'emie universitaire Wallonie-Europe''. This research has made use of the AMBER data reduction package of the Jean-Marie Mariotti Center (\texttt{amdlib v3.0}, available at http://www.jmmc.fr/amberdrs).
\end{acknowledgements}

\bibliographystyle{aa} 
\bibliography{15156} 

\end{document}